\documentclass[aps,prd,onecolumn,groupedaddress]{revtex4-1}

\usepackage{graphicx}
\usepackage{booktabs}
\usepackage{amssymb,bm,mathrsfs,bbm,amscd,epsfig,graphics,color}
\usepackage[tbtags]{amsmath}
\usepackage{lineno}
\usepackage{subfigure}

\begin{document}

\title{ Constraints on branes' geometry from massive vector KK modes}
\author{Chun-E Fu$^1$}\email[E-mail:]{fuche13@mail.xjtu.edu.cn}
\author{Heng Guo $^{2}$} \email[E-mail:]{hguo@xidian.edu.cn}
\author{Ye-Hao Yang$^{1}$} \email[E-mail:]{yyh20000404@stu.xjtu.edu.cn}

\affiliation
{$^1$Institute of Theoretical Physics, School of Physics, Xi'an Jiaotong University, Xi'an 710049, P. R. China\\
$^2$ School of Physics, Xidian University, Xi'an 710071, P. R. China}

\begin{abstract}
  We examine the gauge invariance of massive vector Kaluza-Klein (KK) modes within various 6D brane models. Our analysis reveals that additional constraints on the brane's geometry are essential to maintain the gauge invariance of the massive vector KK modes. However, these conditions are not universally satisfied by brane solutions, leading to loss of gauge invariance. In instances where the brane solutions align with the conditions, we compute the mass spectra of both vector and scalar KK modes, and find some resonances for the KK modes in one of the brane models. Our findings indicate that the presence of a single type of massive scalar KK mode will break the gauge invariance.

\end{abstract}

\maketitle

\section{Introduction}

The concept of extra dimensions and brane worlds has been a topic of interest for many years. Initially, the extra dimensions are supposed to be compactified on the scale of the Planck length \cite{kk,ArkaniHamed1998rs,Antoniadis1998}. While  in the Randall-Sundrum brane model \cite{Randall1999a,Randall1999b}, the extra dimensions were considered infinite, enhancing the possibility of their experimental detection. Our study focuses on the physical phenomena that arise from these extra dimensions. For example, a massless field in the bulk can acquire mass through these extra dimensions, manifesting as a series of Kaluza-Klein (KK) modes in the brane \cite{Karch:2000ct,Youm:2001dt,Ichinose:2002kg,PhysRevD.66.024024,LocalizationPRDLanglois2003,Mukhopadhyaya2004,Melfo:2006hh,Davies:2007tq,Guo:2008ia,Liu2007WeylVo,Koley:2008tn,QformThickRS,Cruz2010zz,Yang2012,Das2012yz,Jones:2013ofa,Chakraborty:2014xda,Zhao2014iqa,Sousa:2014dpa,VaqueraAraujo2014tia,liu2019,Chen_2021,Li_2023,Tan_2022,Tan_2023,tan2023characteristic,wan2023localization,Zhong_2022}. This has profound implications for the $U(1)$ gauge field.

It is well known that the Stueckelberg mechanism \cite{Ruegg:2003ps} or Higgs mechanism \cite{PhysRevLett.13.508} is necessary to preserve the gauge invariance of massive gauge field. These two mechanisms respectively introduce a Stueckelberg field or a Goldstone boson to maintain the symmetry of the gauge field. We observe that for a higher-dimensional massless $U(1)$ gauge field, there are two types of KK modes present in the brane: vector and scalar. Due to the scalar KK modes, these massive vector modes are formulated gauge invariant \cite{PhysRevD.93.064007,Fu2016plb}. We have investigated how the number of extra dimensions, the dimensionality of the brane, and the coupling between the bulk field and other scalar fields, such as the dilaton field, affect this gauge invariance \cite{Fu_2019,Fu_2020,chune2022}.

Our previous discussions were grounded on a conformal metric.
Brane models with one extra dimension under the conformal metric are typically solvable; however, this is not the case for branes with more extra dimensions. For example, many 6D brane models have been constructed with metrics that differ from the conformal metric. Naturally, one might inquire whether the gauge invariance of massive vector KK modes is still remained in these various brane models, and how it relates with the geometry of the branes.

To address this, we  begin by reviewing the method for studying the gauge invariance of the effective action of a bulk free U(1)gauge field. From this, we show that the relationships between the coupling coefficients will determine the gauge invariance. Next, we consider a brane model with the following general line-element:
\begin{equation}
  ds^2=\text{e}^{2B_1(y,z)}\hat{g}_{\mu\nu} dx^\mu\ dx^\nu
  +\text{e}^{2B_2(y,z)} dz^2 +\text{e}^{2B_3(y,z)} dy^2.\label{Genralline-element1}
\end{equation}
Here $\text{e}^{2B_1(y,z)}, \text{e}^{2B_2(y,z)}$ and $\text{e}^{2B_3(y,z)}$ are the functions of two extra dimensions $y$ and $z$. This line-element can be matched with most of the 6D brane models studied in literatures by choosing special $\text{e}^{2B_1(y,z)}, \text{e}^{2B_2(y,z)}$ and $\text{e}^{2B_3(y,z)}$. Since these three factors appear in every coupling coefficients, they impact on the effective action and its gauge invariance. We will demonstrate that certain constraints should be imposed to preserve the gauge invariance of the effective action. Subsequently, we verify whether the brane solutions meet the constraints. Once the requisite conditions are satisfied, we will proceed to calculate the mass spectra of the KK modes. Our findings will demonstrate that when a single type of scalar KK mode is massive, the gauge invariance is disrupted.

Our paper is structured as follows: In Sec. \ref{localizationmechanism}, we review the derivation of the effective action for a massless bulk $U(1)$ field, and point out the key for the gauge invariance of massive vector KK modes. Subsequently, we will discuss the gauge invariance in the three distinct types of brane models, as characterized by the metrics \eqref{case1lineelement1}$\sim$\eqref{case1lineelement3} in Sec. \ref{restritions}. Finally, we summarize our findings and give another way to the constraints in Sec. \ref{summary}.

\section{The key to Gauge Invariance of Massive Vector Kaluza-Klein Modes}\label{localizationmechanism}

We consider a massless  $U(1)$ gauge field $X_{M}(M=0,1,\dots 5)$ as a  perturbation in the bulk, embedding a brane with codimension-2. The action for the field is given by
\begin{eqnarray}\label{bulkaction}
S = -\frac{1}{4} \int d^6 x \sqrt{-g} \; Y^{M_1 M_2} Y_{M_1 M_2},
\end{eqnarray}
where the field strength is defined as $Y_{M_1M_2}=\frac{1}{2}\big(\partial_{M_1}X_{M_2}-\partial_{M_2}X_{M_1}\big)$. In the brane, this field manifests as a series of KK modes. By developing a KK decomposition for the bulk field, we can derive the effective action for these KK modes. The general KK decomposition is chosen as follows:
\begin{subequations}\label{kkdecomp}
\begin{eqnarray}
X_{\mu}(x^{\mu}, y, z) &=& \sum_n \hat{X}_{\mu}^{(n)}(x^{\mu}) \; W_1^{(n)}(y, z), \label{kk1} \\
X_{z}(x^{\mu}, y, z) &=& \sum_n \phi^{(n)}(x^{\mu}) \; W_2^{(n)}(y, z), \label{kk2} \\
X_{y}(x^{\mu}, y, z) &=& \sum_n \varphi^{(n)}(x^{\mu}) \; W_3^{(n)}(y, z), \label{kk3}
\end{eqnarray}
\end{subequations}
where $W_1^{(n)}(y, z)$, $W_2^{(n)}(y, z)$, and $W_3^{(n)}(y, z)$ are functions dependent only on the extra dimensions  $y, z$. The term $\hat{X}_{\mu}^{(n)}(x^{\mu})$ represents the
 $n-$ level vector KK mode, while $\phi^{(n)}(x^{\mu})$ and $\varphi^{(n)}(x^{\mu})$  denote two distinct types of scalar modes. Substituting the KK decomposition \eqref{kkdecomp} into the bulk action \eqref{bulkaction} yields the effective action:
\begin{eqnarray}
S &=& -\frac{1}{4} \sum_{n,n'} \int d^4 x \sqrt{-\hat{g}} \; \bigg[ I_1^{(nn')} \hat{Y}^{(n)}_{\mu\nu} \hat{Y}^{\mu\nu(n')} + \big(I_2^{(nn')} + I_4^{(nn')}\big) \hat{X}^{(n)}_\mu \hat{X}^{\mu(n')} \nonumber \\
&& + I_3^{(nn')} \partial_\mu \phi^{(n)} \partial^\mu \phi^{(n')} - I_6^{(nn')} \big(\partial_\mu \phi^{(n)} \hat{X}^{\mu(n')} + \hat{X}^{(n)}_\mu \partial^\mu \phi^{(n')}\big) \nonumber \\
&& + I_5^{(nn')} \partial_\mu \varphi^{(n)} \partial^\mu \varphi^{(n')} - I_8^{(nn')} \big(\partial_\mu \varphi^{(n)} \hat{X}^{\mu(n')} + \hat{X}^{(n)}_\mu \partial^\mu \varphi^{(n')}\big) \nonumber \\
&& + I_7^{(nn')} \phi^{(n)} \phi^{(n')} + I_9^{(nn')} \varphi^{(n)} \varphi^{(n')} - I_{10}^{(nn')} \big(\phi^{(n)} \varphi^{(n')} + \varphi^{(n)} \phi^{(n')}\big) \bigg]. \label{effectiveAction}
\end{eqnarray}
Here, the coupling constants $I_{1}^{(nn')}\sim I_{10}^{(nn')} $ are expressed as integrals over the extra dimensions. We have not yet provided precise definitions for these integrals, as the metric, which is essential for their calculation, has not been specified. The constants
$I_{2}^{(nn')}$ and $I_{4}^{(nn')}$ are the masses attributed to the vector KK modes, with each part's mass originating from one of the two extra dimensions.  $I_{7}^{(nn')}$ and $I_{9}^{(nn')}$ represent the masses of the two types of scalar modes, respectively. The interaction between the vector and scalar KK modes is denoted by $I_{6}^{(nn')}$ and $I_{8}^{(nn')}$, which we refer to as vector-scalar coupling constants. The masses of the two types of scalar KK modes denote by $I_{7}^{(nn')}$ and $I_{9}^{(nn')}$. They also interact with scalar-scalar coupling $I_{10}^{(nn')}$.  Moreover, the orthogonality conditions, given by
$I_{1}^{(nn')}=\delta_{nn'}, I_{3}^{(nn')}=I_{5}^{(nn')}=2\delta_{nn'}$ serve as fundamental postulates within the framework of this study.

The key to gauge invariance of the effective action is the interrelation between masses of the KK modes with the vector-scalar and scalar-scalar coupling constants. Building on our prior research \cite{PhysRevD.93.064007,Fu2016plb,Fu_2020,Fu_2019,chune2021,chune2022}, we found that the effective action is gauge invariant in branes described by a conformal metric. However, the existing literature presents solutions for 6D branes with non-conformal metrics. Our research aims to extend the investigation of gauge invariance to these 6D brane models. In this scenario, the metric (the geometry) of the brane is different, all the coupling constants will change. Therefore, we will first reconsider whether the gauge invariance is maintained under such variations.

To this end, we derive the equations of motion (EOM) for the KK modes using two distinct approaches. Firstly, the EOM can be directly derived from the effective action \eqref{effectiveAction}:
\begin{subequations}\label{eomfromaction}
\begin{eqnarray}
&& \frac{1}{\sqrt{-\hat{g}}} \partial_{\mu} (I_1^{(nn')} \sqrt{-\hat{g}} \hat{Y}^{\mu\nu(n')}) - \bar{I}^{(nn')} \hat{X}^{\nu(n')} + I_6^{(nn')} \partial^\nu \phi^{(n')} + I_8^{(nn')} \partial^\nu \varphi^{(n')} = 0, \label{effequ1} \\
&& \frac{1}{\sqrt{-\hat{g}}} \partial_{\mu} (I_3^{(nn')} \sqrt{-\hat{g}} \partial^{\mu} \phi^{(n')} - I_6^{(nn')} \sqrt{-\hat{g}} \hat{X}^{\mu(n')}) - I_7^{(nn')} \phi^{(n')} + I_{10}^{(nn')} \varphi^{(n')} = 0, \label{effequ2} \\
&& \frac{1}{\sqrt{-\hat{g}}} \partial_{\mu} (I_5^{(nn')} \sqrt{-\hat{g}} \partial^{\mu} \varphi^{(n')} - I_8^{(nn')} \sqrt{-\hat{g}} \hat{X}^{\mu(n')}) - I_9\;\phi^{(n)}
  -\lambda_{10}\;\varphi^{(n)}=0,
\end{eqnarray}
\end{subequations}
with $\bar{I}^{(nn')}=I_{2}^{(nn')}+I_{4}^{(nn')}$. On the other hand, substituting the KK decomposition into the EOM of the bulk field $\partial_{M}(\sqrt{-g}Y^{MN})=0$, we have:
\begin{subequations}\label{eomfrombulkeom}
\begin{eqnarray}\label{effequ11}
  &&\!\!\!\!\!\!\!\!\frac{1}{\sqrt{-\hat{g}}}\;\partial_{\mu_1}
  (\sqrt{-\hat{g}}\;\hat{Y}^{\mu_1\mu_2(n)})\;
  -(\lambda_1+\lambda_2)\;\hat{X}^{\mu_2(n)}
  +\lambda_3\;\partial^{\mu_2}\varphi^{(n)}
  +\lambda_4\;\partial^{\mu_2}\phi^{(n)}=0 ,\\
  \label{effequ22}
 &&\frac{1}{\sqrt{-\hat{g}}}\partial_{\mu_1}
  (\sqrt{-\hat{g}}\;\partial^{\mu_1}\phi^{(n)}
  -\lambda_5\;\sqrt{-\hat{g}}\;\hat{X}^{\mu_1(n)})
  -\lambda_6\;\phi^{(n)}
  +\lambda_7\;\varphi^{(n)}=0,\\
  \label{effequ33}
 &&\frac{1}{\sqrt{-\hat{g}}}\partial_{\mu_1}
  (\sqrt{-\hat{g}}\;\partial^{\mu_1}\varphi^{(n)}
  -\;\lambda_8\sqrt{-\hat{g}}\;\hat{X}^{\mu_1(n)})
  +\lambda_9\;\phi^{(n)}
  -\lambda_{10}\;\varphi^{(n)}=0,
\end{eqnarray}
\end{subequations}
where $\lambda_1\sim\lambda_{10}$ are defined differently in different branes, which will be shown in the following sections. The two groups of EOM \eqref{eomfromaction} and \eqref{eomfrombulkeom} must be consistent. It is from this consistency that we can deduce the equations that the KK modes satisfy, as well as the interrelations among the coupling constants. Armed with these interrelations, we can then investigate the gauge invariance of the massive vector KK modes. It will be found that additional constraint conditions on the brane geometry are necessary to obtain a gauge-invariant effective action.

\section{Constraints on the geometry from the gauge invariance }\label{restritions}

With the metric \eqref{Genralline-element1} and the KK decomposition \eqref{kkdecomp}, the parameters  $\lambda_1\sim\lambda_{10}$ featured in the EOM \eqref{eomfrombulkeom} are explicitly defined as:
\begin{eqnarray}
\lambda_1 &\equiv& \frac{\partial_z
\big(\text{e}^{2B_1-B_2+B_3}\;\partial_z W_1^{(n)}\big)\text{e}^{-(B_2+B_3)}}
  {2W_1^{(n)}}, ~~
\lambda_2 \equiv \frac{\partial_y
\big(\text{e}^{2B_1+B_2-B_3}\;\partial_y W_1^{(n)}\big)\text{e}^{-(B_2+B_3)}}
  {2W_1^{(n)}},\nonumber\\
\lambda_3 &\equiv& \frac{\partial_y \big(\text{e}^{2B_1+B_2-B_3}W^{(n)}_3\big)
\text{e}^{-(B_2+B_3)}
  }{2W_1^{(n)}},~~~~~
  \lambda_4\equiv \frac{\partial_z \big(\text{e}^{2B_1-B_2+B_3}W^{(n)}_2\big)
  \text{e}^{-(B_2+B_3)}
  }{2W_1^{(n)}},\nonumber\\
\lambda_5&\equiv&\frac{
  \partial_zW_1^{(n)}}{W_2^{(n)}},
\lambda_6 \equiv \frac{
  \partial_y\big(\text{e}^{4B_1-B_2-B_3}\partial_y
  W_2^{(n)}\big)\text{e}^{-(2B_1-B_2+B_3)}
  }{W_2^{(n)}},
\lambda_7\equiv\frac{\partial_y
\big(\text{e}^{4B_1-B_2-B_3}\partial_z W_3^{(n)}
  \big)\text{e}^{-(2B_1-B_2+B_3)}}{W_2^{(n)}},\nonumber\\
\lambda_8 &\equiv & \frac{
  \partial_yW_1^{(n)}}{W_3^{(n)}},
\lambda_9 \equiv \frac{\partial_z
\big(\text{e}^{4B_1-B_2-B_3}\partial_y W_2^{(n)}
  \big)\text{e}^{-(2B_1+B_2-B_3)}}{W_3^{(n)}},
\lambda_{10}\equiv \frac{
  \partial_z\big(\text{e}^{4B_1-B_2-B_3}\partial_z W_3^{(n)}\big)
\text{e}^{-(2B_1+B_2-B_3)}}{W_3^{(n)}}.\nonumber
\end{eqnarray}

For convenience, we introduce the constants $C_{1}^{(n)}, C_{2}^{(n)}, C_{3}^{(n)} $ to denote the vector-scalar and scalar-scalar couplings, and  $m_{1}^{(n)},m_{2}^{(n)},m_{\phi}^{(n)},m_{\varphi}^{(n)}$ to represent the masses of the KK modes:
\begin{eqnarray}
&&I_{2}^{(nn)}=\frac{1}{2}\; {m_{1}^{(n)}}^2,~
  I_{4}^{(nn)}=\frac{1}{2}\; {m_{2}^{(n)}}^{2},~
  I_{7}^{(nn)}=2\; {m_\phi^{(n)}}^{2},~
  I_{9}^{(nn)}=2\; {m_\varphi^{(n)}}^{2},\\
&&I_{6}^{(nn')}=C_{1}^{(n)}\;\delta_{n'n},~~
I_{8}^{(nn')}=C_{2}^{(n)}\;\delta_{n'n},~~
I_{10}^{(nn')}=C_{3}^{(n)}\;\delta_{n'n}.
\end{eqnarray}
 By comparing Eq. \eqref{eomfromaction} with Eq. \eqref{eomfrombulkeom}, we derive the following equations:
\begin{eqnarray}
-\partial_z
\big(\text{e}^{2B_1-B_2+B_3}\;\partial_z W_1^{(n)}\big)\text{e}^{-B_2-B_3}&=& m_1^{(n)2}W^{(n)}_1,\label{Schro1}\\
-{\partial_y
\big(\text{e}^{2B_1+B_2-B_3}\;\partial_y W_1^{(n)}\big)\text{e}^{-B_2-B_3}}&=& m_2^{(n)2}W^{(n)}_1,\label{Schro2}\\
-{\partial_y\big(\text{e}^{4B_1-B_2-B_3}\partial_y
  (W_2^{(n)})\big)
  \text{e}^{-(2B_1-B_2+B_3)}}&=& m_\phi^{(n)2}W^{(n)}_2,\label{Schro3}\\
-{\partial_z\big(\text{e}^{4B_1-B_2-B_3}\partial_z (W_3^{(n)})\big)
\text{e}^{-(2B_1+B_2-B_3)}}&=& m_\varphi^{(n)2}W^{(n)}_3.\label{Schro4}
\end{eqnarray}
And
\begin{eqnarray}
 -2\;C_{1}^{(n)}\;{W_1^{(n)}}&=&{\partial_z (\text{e}^{2B_1-B_2+B_3}W^{(n)}_2)
  \;\text{e}^{-B_2-B_3}},\label{Acoupl1}\\
 \frac{1}{2}\;C_{1}^{(n)}\;{W_2^{(n)}}&=&
  \partial_z\big(W_1^{(n)}\big),\label{Acoupl2}\\
 -2\;C_{2}^{(n)}\;{W_1^{(n)}}&=&{\partial_y (\text{e}^{2B_1+B_2-B_3}W^{(n)}_3)
  \;\text{e}^{-B_2-B_3}},\label{Acoupl3}\\
 \frac{1}{2}\;C_{2}^{(n)}\;{W_3^{(n)}}&=&
  \partial_y\big(W_1^{(n)}\big),\label{Acoupl4}\\
 -\frac{1}{2}\;C_{3}^{(n)}\;{W_2^{(n)}}&=&{\partial_y
\big(\text{e}^{4B_1-B_2-B_3}\partial_z (W_3^{(n)})
  \big)\text{e}^{-(2B_1-B_2+B_3)}},\label{Acoupl5}\\
 -\frac{1}{2}\;C_{3}^{(n)}\;{W_3^{(n)}}&=&{\partial_z
\big(\text{e}^{4B_1-B_2-B_3}\partial_y (W_2^{(n)})
  \big)\text{e}^{-(2B_1+B_2-B_3)}}.
  \label{Acoupl6}
\end{eqnarray}
Then we can establish the following relationships between the coupling constants.
\begin{itemize}
\item
Firstly, by differentiating with respect to $z$ to Eq. \eqref{Acoupl1}, we derive  a relationship between $C_{1}^{(n)}$ and $m_1^{(n)}$ with the aid of Eqs. \eqref{Acoupl2} and \eqref{Schro1}. Similarly, a relationship between $C_{2}^{(n)}$ and $m_2^{(n)}$ is also established. These relationships are presented as follows:
\begin{equation}
C_{1}^{(n)2}=m_1^{(n)2},~~C_{2}^{(n)2}=m_2^{(n)2}.\label{c12vector}
\end{equation}
\item
Secondly, we try to find the relationship between $C_{1}^{(n)},C_{2}^{(n)}$ and $C_{3}^{(n)}$. By differentiating with respect to $y$ to Eq. \eqref{Acoupl1} and considering Eq. \eqref{Acoupl4}, there is an equation about $W_3^{(n)}$ and $W_2^{(n)}$:
\begin{equation}
-\frac{1}{2}\;C_{3}^{(n)}\;{W_2^{(n)}}=\frac{C_{3}^{(n)}}{2C_{1}^{(n)}C_{2}^{(n)}}\;
\partial_z\big({\partial_y (\text{e}^{2B_1+B_2-B_3}W^{(n)}_3)
  \;\text{e}^{-B_2-B_3}}\big),
\end{equation}
which must be consistent with Eq. \eqref{Acoupl6}. From this consistency, we confirm that a constraint condition on the geometry of the brane, given by
\begin{eqnarray}\label{condition1}
\partial_{y,z} ({2B_1+B_2-B_3})&=&0,\label{Byz1}\\
\partial_z(B_1-B_3)=0,~~\partial_y(B_1-B_2)&=&0,\label{B123}
\end{eqnarray}
must be introduced to establish a relationship between $C_{1}^{(n)}$, $C_{2}^{(n)}$, and $C_{3}^{(n)}$:
\begin{equation}
C_{3}^{(n)2}=2C_{1}^{(n)}C_{2}^{(n)}.\label{c123}
\end{equation}
This relationship, Eq. \eqref{c123}, is crucial to the gauge invariance of the effective action. In contrast, for scenarios involving branes with a single extra dimension, no such constraint on the brane is required.

At the same time, with the Eqs. \eqref{Acoupl2}, \eqref{Acoupl3} and \eqref{Acoupl5}, one more condition is found:
\begin{eqnarray}\label{condition1}
\partial_{y,z} ({2B_1-B_2+B_3})=0.\label{Byz2}
\end{eqnarray}
Finally, by considering all the constrains from Eqs.\eqref{Byz1}, \eqref{c123} and \eqref{Byz2}, we derive the final constrains on the geometry in 6D brane world with line-element \eqref{Genralline-element1}:
\begin{eqnarray}\label{conditionfinal}
\partial_{y,z} B_1=0,~~\partial_{y,z}(B_2-B_3)=0,~~
\partial_zB_1=\partial_zB_3,~~\partial_yB_1=\partial_yB_2.
\end{eqnarray}

\item Furthermore, by taking the derivative with respect to $y$ of Eq. \eqref{Acoupl2} and with respect to $z$ of Eq. \eqref{Acoupl4}, we show that $C_{2}^{(n)}\partial_z W_3^{(n)}=C_{1}^{(n)}\partial_y W_2^{(n)}$. This relationship allows Eq. \eqref{Acoupl5} to be equivalent to Eq. \eqref{Schro3}, thereby establishing a connection between $C_{2}^{(n)}$ and $m_\phi^{(n)}$. A similar relationship between $C_{1}^{(n)}$ and $m_\varphi^{(n)}$ can also be derived. These two relationships are expressed as:
\begin{equation}
C_{2}^{(n)2}=m_\phi^{(n)2},~~~C_{1}^{(n)2}=m_\varphi^{(n)2}.\label{c12scalar}
\end{equation}
\end{itemize}

Using the relationships \eqref{c12vector}, \eqref{c123}, and \eqref{c12scalar}, the effective action, as given by Eq. \eqref{effectiveAction}, can be rewritten as:
\begin{eqnarray}\label{gaugeinvareff}
S_{\text{eff}}&=&-\frac{1}{4}\sum_n\int d^{4}x \sqrt{-\hat{g}} \bigg[ \hat{Y}^{(n)}_{\mu_1\mu_2}\;\hat{Y}^{(n)\mu_1\mu_2} - \frac{1}{2}\sum_n \big(\partial_{\mu}\phi^{(n)}-\frac{1}{2}\;C_1^{(n)}\;\hat{X}_{\mu}^{(n)}\big)^2 \nonumber \\
&& - \frac{1}{2}\sum_n \big(\partial_{\mu}\varphi^{(n)}-\frac{1}{2}\;C_2^{(n)}\;\hat{X}_{\mu}^{(n)}\big)^2 - \frac{1}{2}\sum_n\big(C_2^{(n)}\;\phi^{(n)}-C_1^{(n)}\;\varphi^{(n)}\big)^2 \bigg].
\end{eqnarray}
This formulation maintains gauge invariance under the transformations:
\begin{align*}
\hat{X}_\mu &\rightarrow \hat{X}_\mu + \partial_\mu \gamma^{(n)}, \\
\phi^{(n)} &\rightarrow \phi^{(n)} + C_1^{(n)} \gamma^{(n)}, \\
\varphi^{(n)} &\rightarrow \varphi^{(n)} + C_2^{(n)} \gamma^{(n)},
\end{align*}
with $\gamma^{(n)}$ being a scalar field. Despite variations in the definitions of the coupling constants compared to those in branes with a conformal metric, the structure of the effective action, Eq. \eqref{gaugeinvareff}, remains unchanged. Notably, the geometric constraint, Eq. \eqref{conditionfinal}, is introduced here and is essential for achieving this gauge-invariant effective action. Moreover, we can check various brane models, in which the gauge invariance of the massive vector KK modes can be preserved. In the following, we give some examples.

In the following, we will examine three cases based to 6D brane models in former literatures, and check whether the constrain conditions \eqref{conditionfinal} are satisfied. If the conditions are met, we will  proceed to calculate the masses of the KK modes.

\subsection{case I: $B_1(y,z)=B_1(z), B_2(y,z)=B_3(y,z)=B_2(z)$}

For most of the 6D brane models, to solve the solutions the extra dimensions are assumed as only the functions of one of the extra dimensions $B_1(y,z)=B_1(z)$. In references \cite{Gogberashvili2004,Aguilar2006,Dzhunshaliev2008}, the authors have discussed solutions for a brane with codimension-2, as described by the line-element given as:
\begin{equation}
ds^2=\phi^{2}(r)\hat{g}_{\mu\nu} dx^\mu\ dx^\nu
  -\lambda(r)(dr^2 +r^2d\theta^2),\label{case1lineelement1}
\end{equation}
 For this line-element, the additional spatial part is conformally equivalent to a Euclidean space. The metric function $\lambda(r)$ associated with the extra dimension $r$, serves as a conformal factor for the two-dimensional Euclidean metric.

It is easy to find that the conditions \eqref{conditionfinal} become:
\begin{eqnarray}\label{restr1}
r^{-1}=\frac{\phi'}{\phi}-\frac{\lambda'}{2\lambda}.
\end{eqnarray}
In the referenced literature \cite{Gogberashvili2004,Aguilar2006,Dzhunshaliev2008}, a brane solution is provided:
\begin{eqnarray}\label{branesol1}
\phi(r) = \frac{c^b + a r^b}{c^b + r^b}, \quad \lambda(r) = \frac{\rho^2 \phi'}{r},
\end{eqnarray}
where $a$ (with $a > 1$), $b$, and $c$ are constants, and $\rho$ is an integration constant with units of length. We find that this solution, Eq. \eqref{branesol1}, does not satisfy the condition \eqref{restr1} except when $r = \frac{c}{a}$. This suggests that in this brane model, the gauge invariance of the massive vector Kaluza-Klein modes is only realized at a fixed value of $r$, specifically $r = \frac{c}{a}$.

However, according to Eqs. \eqref{Schro1}, when $r$ is constant, the vector Kaluza-Klein modes and the scalar modes $\phi^{(n)}$ cannot obtain masses from this extra dimension. In contrast, the vector mode and another scalar mode $\varphi^{(n)}$ could gain masses from another compactified extra dimension $\theta$. Consequently, the effective action transforms into:
\begin{equation}\label{brokenaction}
S_{eff} = -\frac{1}{4} \sum_n \int d^4 x \sqrt{-\hat{g}} \left[ \hat{Y}^{(n)}_{\mu\nu} \hat{Y}^{\nu\mu(n)} - 2 \left( \partial_\mu \phi^{(n)} - \frac{1}{2} C_1^{(n)} \hat{X}^{(n)}_\mu \right)^2 - 2 C_1^{(n)2} \varphi^{(n)2} \right].
\end{equation}
It becomes evident that the gauge invariance of this effective action is broken.

\subsection{case II: $B_1(y,z)=B_1(y), B_2(y,z)=B_2(y),~B_3(y,z)=0$}

 In the referenced works \cite{Oda2000zj,Gregory2000,Gherghetta2000,Cline2003,Koley:2006bh,Sousa:2014dpa}, the authors have investigated a brane model described by the line-element:
\begin{equation}
  ds^2=\text{e}^{-A(y)}\hat{g}_{\mu\nu} dx^\mu\ dx^\nu
+\text{e}^{-B(y)}dz^2+dy^2,\label{case1lineelement2}
\end{equation}
where the extra dimension is characterized as a 1-sphere with the range $0 \leq y < \infty$ and $0 \leq z \leq 2\pi$. Comparing the general metric \eqref{Genralline-element1}, we have $B_1=-\frac{1}{2}A(y), B_2=-\frac{1}{2}B(y)$. For this case, the imposing constraint on the geometry \eqref{conditionfinal} becomes:
\begin{equation}\label{restr2}
\partial_{y}A = \partial_{y}B.
\end{equation}
Despite the definitions of the constants $C_{1}^{(n)}$ to $C_{3}^{(n)}$ differing in this context, the gauge-invariant effective action, as given by Eq. \eqref{gaugeinvareff}, can still be formulated under the constraint condition \eqref{restr2}. We discover that there exist two distinct brane solutions that meet this criterion.

Prior to presenting these brane solutions, we aim to simplify Eqs. \eqref{Schro1} to \eqref{Schro4}, which now are equal to:
 \begin{eqnarray}
-\partial_y
  \big(e^{-A-\frac{1}{2}B}
  \partial_y W^{(n)}_1
  \big)\;e^{\frac{1}{2}B}&=& m_1^{(n)2}W^{(n)}_1,\label{case2sch1}\\
-\partial_z
  \big(e^{-A+\frac{1}{2}B}
  \partial_z W^{(n)}_1
  \big)\;e^{\frac{1}{2}B}&=& m_2^{(n)2}W^{(n)}_1,\label{case2sch2}\\
  -\partial_y\big(e^{-2A+\frac{1}{2}B}\partial_y
  W_2^{(n)}\big)
  e^{A-\frac{1}{2}B}&=& m_\phi^{(n)2}W^{(n)}_2,\label{case2sch3}\\
-\partial_z\big(e^{-2A+\frac{1}{2}B}\partial_z W_3^{(n)}\big)
  e^{A+\frac{1}{2}B}&=& m_\varphi^{(n)2}W^{(n)}_3,\label{case2sch4}
\end{eqnarray}
This simplification is instrumental in calculating the mass spectra of the vector and scalar KK modes under the orthogonality conditions. To this end, we carry out a coordinate transformation $dy = e^{-\frac{1}{2}A} d\bar{y}$, and define $W_1^{(n)} = e^{\frac{1}{2}A} \bar{W}_1^{(n)}$ and $W_2^{(n)} = e^{\frac{1}{2}A} \bar{W}_2^{(n)}$. Consequently, Eqs. \eqref{case2sch1} and \eqref{case2sch3} are transformed into a pair of Schrodinger-like equations:
\begin{eqnarray}\label{y1sch}
-\partial_{\bar{y},\bar{y}}\bar{W}_1^{(n)} + V_{\text{eff}} \bar{W}_1^{(n)} &=& m_1^{(n)2} \bar{W}^{(n)}_1, \\
-\partial_{\bar{y},\bar{y}}\bar{W}_2^{(n)} + V_{\text{eff}} \bar{W}_2^{(n)} &=& m_{\phi}^{(n)2} \bar{W}^{(n)}_2 = m_1^{(n)2} \bar{W}^{(n)}_2, \label{y2sch}
\end{eqnarray}
where the effective potential is expressed as
\begin{equation}\label{veff}
 V_{\text{eff}} = \frac{1}{4} (\partial_{\bar{y}}A)^2 - \frac{1}{2} \partial_{\bar{y},\bar{y}}A.
\end{equation}
Additionally, with the transformation $dz = e^{-\frac{1}{2}(A-B)} d\bar{z}$ and $W_3^{(n)} = e^{\frac{1}{2}A} \bar{W}_3^{(n)}$, Eqs. \eqref{case2sch2} and \eqref{case2sch4} are reformulated as:
\begin{eqnarray}\label{z1sch}
\partial_{z,z}\bar{W}^{(n)}_1 = m_2^{(n)2} \bar{W}^{(n)}_1, \quad \partial_{z,z}\bar{W}^{(n)}_3 = m_{\varphi}^{(n)3} \bar{W}^{(n)}_3 = m_2^{(n)2} \bar{W}^{(n)}_3. \label{z2sch}
\end{eqnarray}

Given that the extra dimension $z$ is compact, it is evident from Eq. \eqref{z1sch} that the vector KK modes can acquire a portion of their mass from this dimension, whereas the scalar modes $\varphi^{(n)}$ obtain their full mass from $z$. The ability for the vector modes and the scalar modes $\phi^{(n)}$ to acquire mass from the another dimension $y$ is determined by the effective potential $V_{\text{eff}}$, which depends on the brane's warp factor. We proceed to present two types of brane solutions that fulfill the condition \eqref{restr2}, accompanied by the respective shapes of $V_{\text{eff}}$.
\begin{itemize}
\item In Refs.\cite{Oda2000zj,Gregory2000,Gherghetta2000} there is a brane solution:
\begin{eqnarray}\label{case2warpfactor1}
A(y)=cy,~~B(y)=cy-2\log R_0,
\end{eqnarray}
where $c$ is a positive constant, and $R_0$ being a length scale. Under the coordinate transformation $dy = e^{-\frac{1}{2}A} d\bar{y}$, the warp factor can be written as $A(\bar{y})=2\ln(\frac{1}{2}c\bar{y})$, then the effective potential \eqref{veff} is derived as $V_{\text{eff}}=\frac{2}{y^2}$. We plot it in Fig.\ref{veff1}. For this case, there is no bound or resonance KK mode obtained from this extra dimension.
\begin{figure*}[htb]
\begin{center}
\includegraphics[width=6cm]{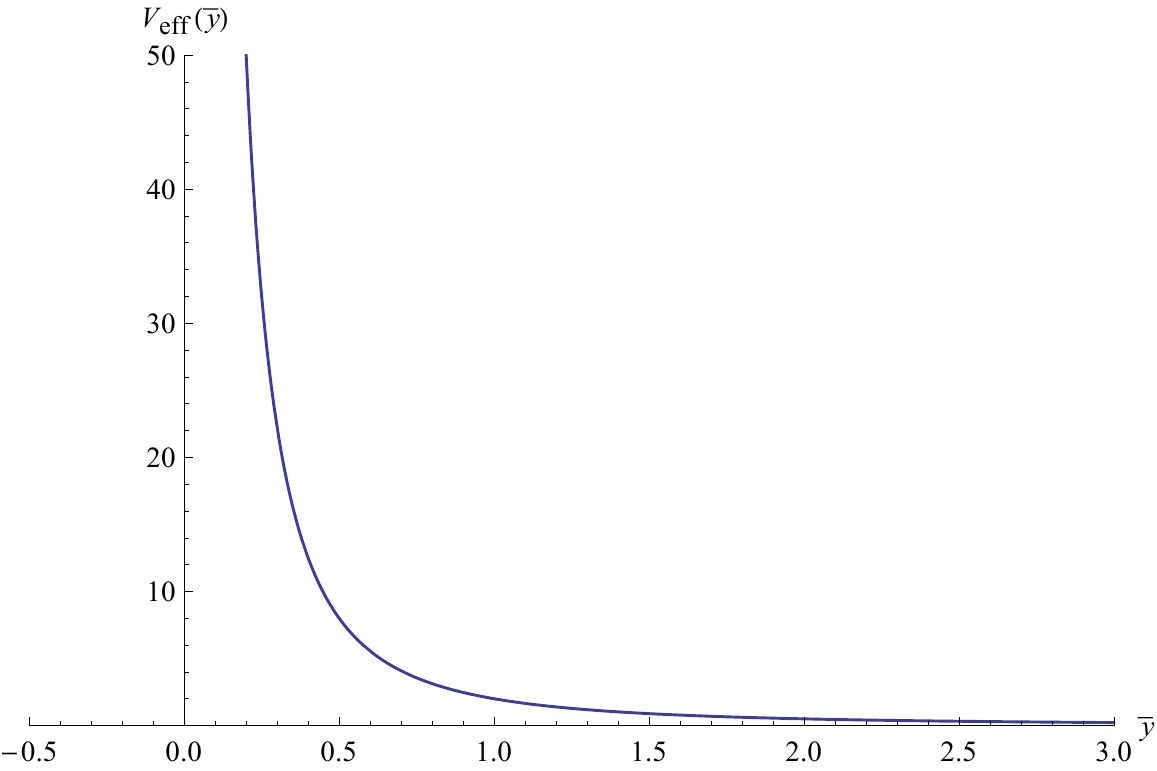}
\end{center}
 \caption{$V_{\text{eff}}$ in brane model with warp factor \eqref{case2warpfactor1} under $c=1$ }
 \label{veff1}
\end{figure*}

\item In Ref.\cite{Sousa:2014dpa,SOUSA2012579}, there is also a solution satisfying \eqref{restr2}:
\begin{equation}\label{case2warpfactor2}
A(y)=B(y)=\beta \ln \cosh (a y)+\frac{\beta}{2} \tanh^2 (a y)
\end{equation}
with $a, \beta(\beta>0)$ two constants. And the effective potential in this brane model is plotted in Fig.\ref{veff2}.
\begin{figure}[htb]
  \centering
  \subfigure[$~~a=1$]{\includegraphics[width=5cm]{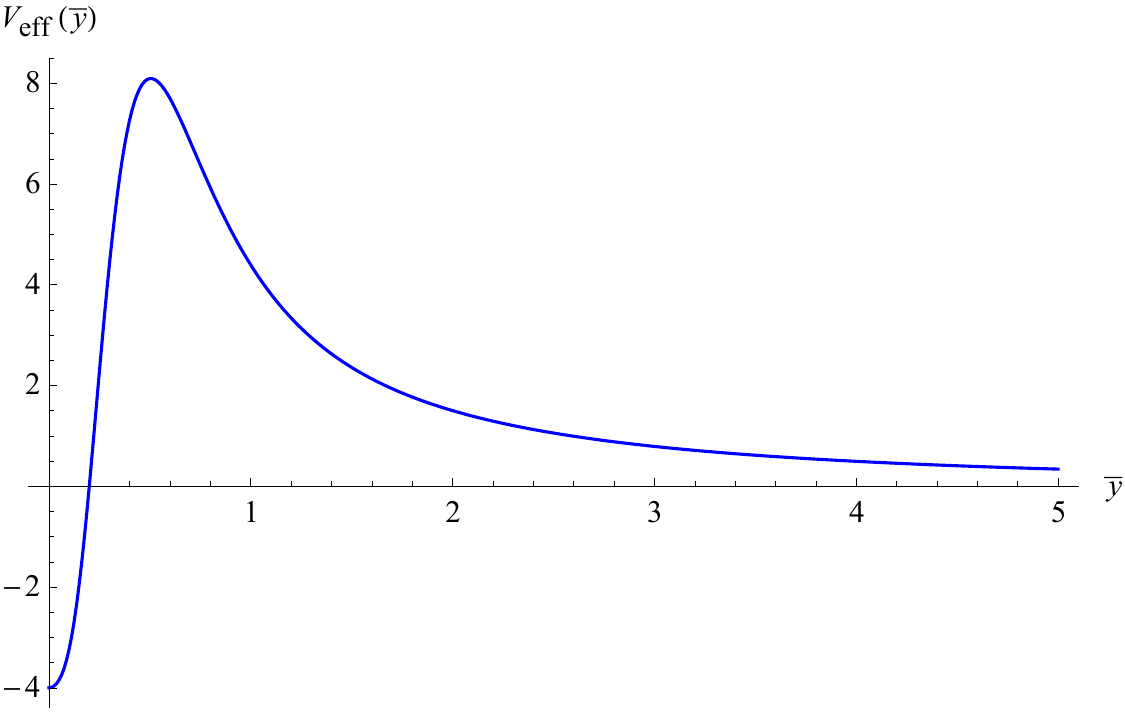}}\hspace{10mm}
  \subfigure[$~~a=2.5$]{\includegraphics[width=5cm]{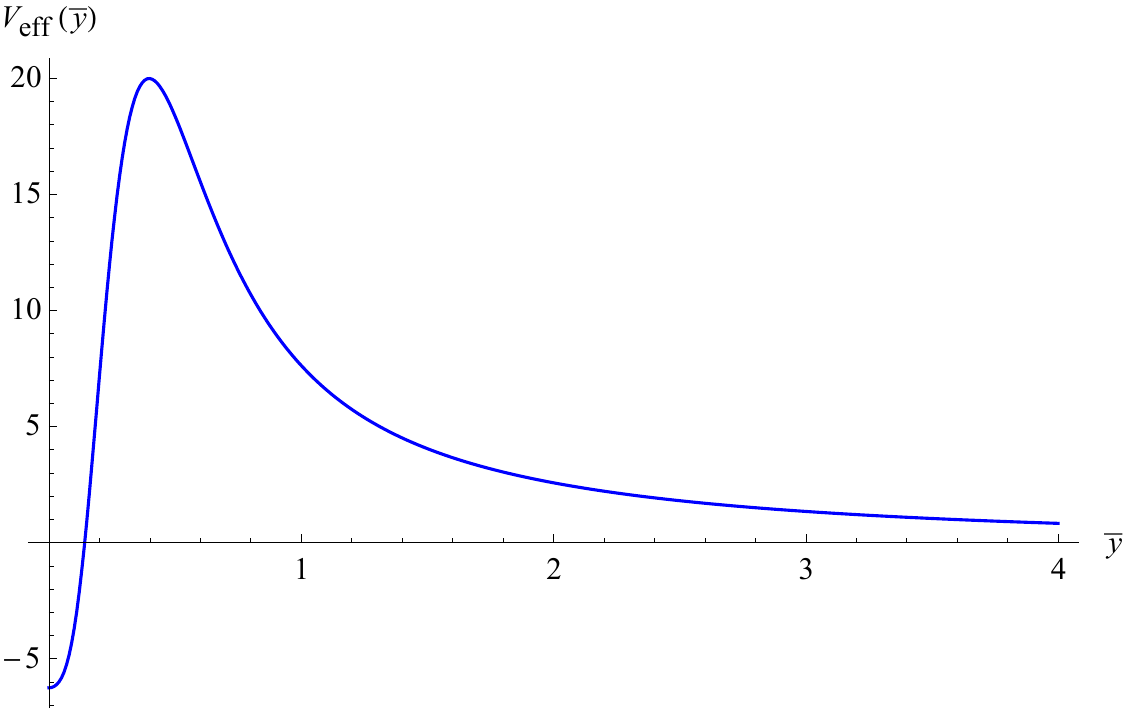}}\hspace{10mm}
  \subfigure[$~~a=3$]{\includegraphics[width=5cm]{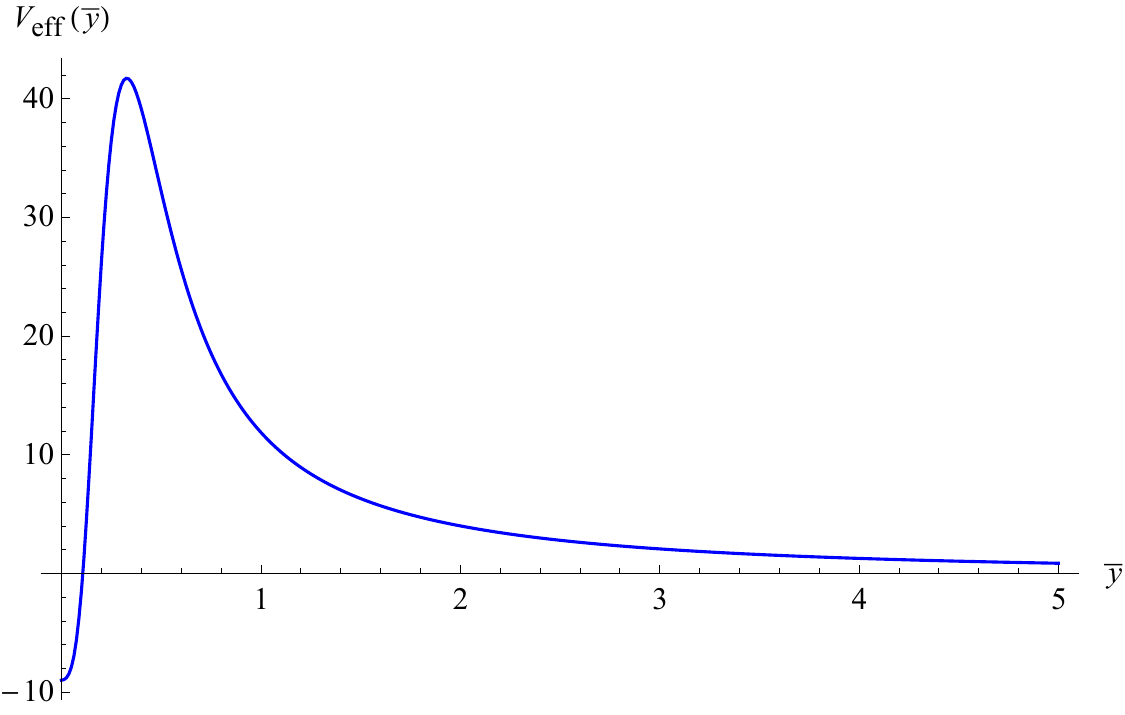}}
  \caption{$V_{\text{eff}}$ in brane model with warp factor \eqref{case2warpfactor2} under different $a$ and $b=1$. }
  \label{veff2}
\end{figure}
Although there is no bound Kaluza-Klein mode, some resonances exist. We present the numerical results using the method outlined in Refs.~\cite{Liu2009,Liu2011,Liu2017}. To identify the resonances, we define a relative ratio:
\begin{eqnarray}
    P(m^2) = \frac{\int_0^{\eta \bar{y}_0} f^{(n)2} d\bar{y}}{\int_0^{\bar{y}_0} f^{(n)2} d\bar{y}},
\end{eqnarray}
where $\eta_0<1$.  Resonances can be identified by the peaks of this relative ratio. The relative ratios for different $a$ are shown in Fig.~\ref{reson}.
\begin{figure*}[htb]
  \centering
  \subfigure[$~~a=1,~~m_1^{(1)2}=3.42068$]{\includegraphics[width=6cm]{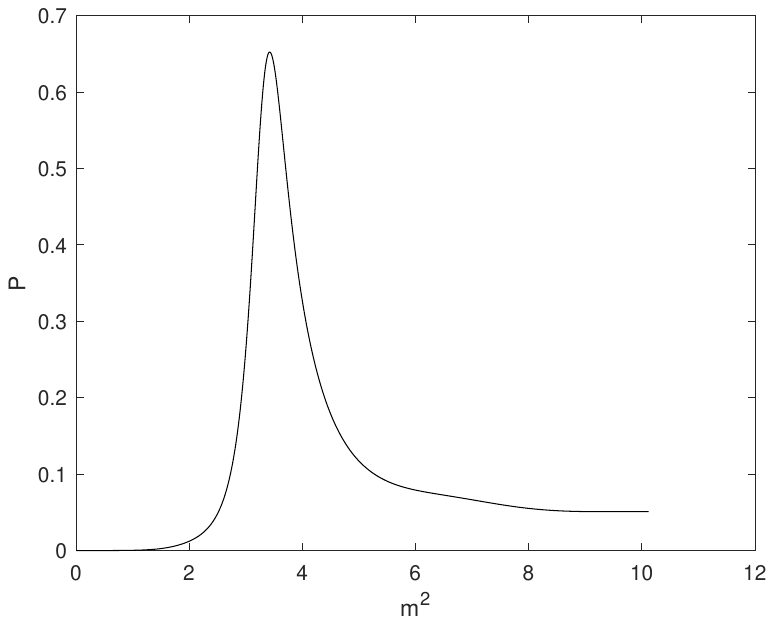}}\hspace{-10mm}
  \subfigure[$~~a=2.5~~m_1^{(1)2}=8.29298$]{\includegraphics[width=6cm]{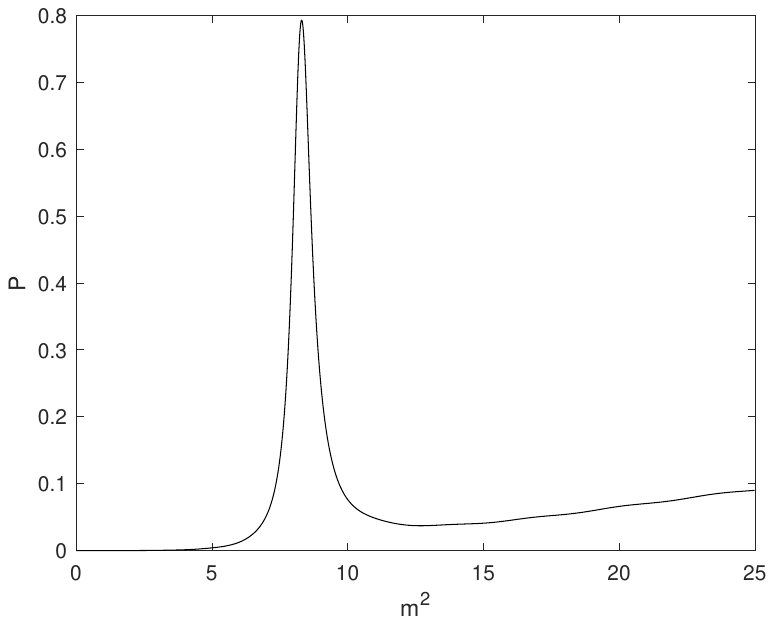}}\hspace{-10mm}
  \subfigure[$~~a=3~~m_1^{(1)2}=16.41179$]{\includegraphics[width=6cm]{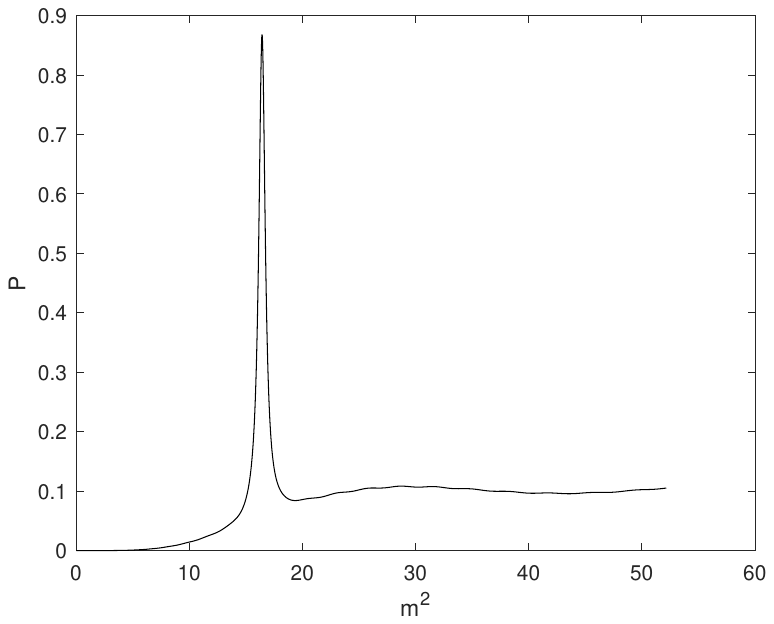}}
  \caption{ Relative ratio denoting the resonances in brane model with warp factor \eqref{case2warpfactor2} under different $a$ and $b=1$. }
  \label{reson}
\end{figure*}
We observe that resonances appear due to the warp factor $A(r)$. Interestingly, for a massive vector KK mode, its mass consists two parts, one of which has a finite lifetime on the brane. Meanwhile, resonances also exist for the scalar KK mode $\varphi^{(n)}$. When they vanish the gauge invariance of the effective action will be destroyed.
\end{itemize}

\subsection{case III: $B_2(y,z)=B_3(y,z)$}
For the two cases discussed earlier, the warp factors are dependent on a single extra dimension. However, it is common for warp factors to be associated with both extra dimensions, as exemplified by Eq. \eqref{case1lineelement3}:
 \begin{equation}
  ds^2=\text{e}^{2B_1(y,z)}\hat{g}_{\mu\nu} dx^\mu\ dx^\nu
  +\text{e}^{2B_2(y,z)} (dy^2 +dz^2).\label{case1lineelement3}
\end{equation}
Branes of this nature are known as multi-warped branes. It is clear that for this case,
 the gauge-invariant effective action needs the constraints on the brane's geometry:
\begin{equation}\label{restr3}
\partial_z B_1 = \partial_z B_2, \quad \partial_y B_1 = \partial_y B_2, \quad \partial_{y,z} B_1 = 0.
\end{equation}
In the referenced literature \cite{PhysRevD.105.024068,PhysRevD.106.084003}, solutions for a brane model characterized by the line-element \eqref{case1lineelement3} have been obtained. However, these solutions do not satisfying these constraints.

\section{Discussion and summary}\label{summary}

This study begun with an examination of the methodology for deriving the effective action of a massless bulk $U(1)$ gauge field through a general KK decomposition within branes of codimension 2. The effective action implies the existence of two distinct types of scalar KK modes that couple with the massive vector modes. While we have established that the effective action maintains gauge invariance in brane models with a conformal metric. However, the solvable 6D branes are usually constructed within the non-conformal metrics. By comparing the EOM for the KK modes (deriving from two ways), we revealed that to preserve the gauge invariance of the effective action in these 6D branes, certain constraints on the brane's geometry must be introduced.

Nevertheless, the gauge invariance can be easily compromised. Firstly, if the brane solutions do not conform to the imposed constraint conditions, the formulation of a gauge-invariant effective action becomes unfeasible. Secondly, even when the brane solutions match the constraints, gauge invariance is only preserved if both types of massive bound scalar KK modes are present within the brane.

Furthermore, from the effective action  \eqref{effectiveAction}, we see that if there are no couplings between different types of scalar KK  modes, which are from the term $Y_{yz}Y^{yz}$ in the bulk action of the field, the effective action resembles that of a 5D brane model. In this case, no additional constraints on the geometry are required to preserve the gauge invariance.  In our another work \cite{fu2024}, we also discover that the constraints are related to the assumption that only the KK modes under the same level can interact with each other.

\section*{Acknowledgement}
This work was supported by the Natural Science Foundation of Shaanxi Province (No. 2022JQ-037). This work was  also supported by the National Natural Science Foundation of China (Grants No. 11305119), the Natural Science Basic Research Plan in Shaanxi Province of China
(No. 23JSQ001, No. 2020JM-198) and the Fundamental Research
Funds for the Central Universities (No. JB170502), and the 111 Project (B17035).

\section{References}
\bibliographystyle{aapmrev4-2}

\bibliography{constraints_new}

\end{document}